\documentclass[12pt]{article}

\usepackage{amsmath}
\usepackage[dvips]{graphicx}
\usepackage{bm}
\usepackage{color} 

\newcommand{\bea}{\begin{eqnarray}}  
\newcommand{\eea}{\end{eqnarray}}  

\baselineskip
\newcommand{\xp}{\ensuremath{x_{I\!\!P}}}

\begin{document}

\begin{flushright}
MAN/HEP/2006/31\\
\end{flushright}

\begin{center}
\vspace*{0cm}

{\Large {\bf Colour dipoles and the HERA data}\footnote{{\bf This talk is based on
    work carried out in collaboration with J.R.~Forshaw and R.~Sandapen, \cite{FS04,fss:06a} }}}\\

\vspace*{1cm}

Graham.~Shaw$^{1}$

\vspace*{0.2cm}
$^{1}$Particle Physics Group,\\Department of Physics and Astronomy,\\
University of Manchester,\\
Manchester. M13 9PL. England.

\end{center}

\begin{abstract}
\noindent
We confront a very wide body of HERA diffractive electroproduction
  data with the predictions of the colour dipole model. In doing so we focus
  upon three  different parameterisations of the dipole scattering
  cross-section, whose parameters are fixed by analysis of DIS structure
  function  data. This analysis  strongly suggests the presence of saturation
  effects,  but is not definitive because  the conclusion requires the
  inclusion of data at low-$Q^2$ values.

Having fixed the parameters of the models from the DIS structure function data,
 the resulting dipole cross sections can be used to make genuine predictions
 for other reactions. Good agreement is obtained for all observables, as is
 illustrated here for deeply virtual Compton scattering(DVCS)  and diffractive
 deep inelastic scattering(DDIS). There can be no doubting the success of the
 dipole scattering approach and more precise observations 
are needed in order to expose its limitations

\end{abstract}


                                                                 


\section{Introduction}

\begin{figure}[tbh]
\begin{center}
\includegraphics[width=6cm]{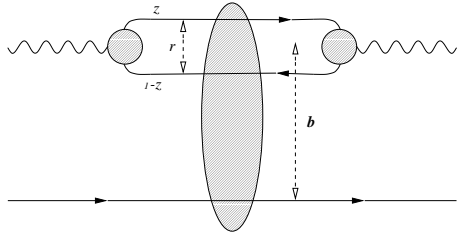}
\end{center}
\caption{The colour dipole model for elastic Compton scattering $\gamma^{\ast
}p\rightarrow\gamma^{\ast}p$.}%
\label{f2dipole}%
\end{figure}

In the colour dipole model ~\cite{NZ91,Mueller94}, the forward amplitude for virtual Compton
scattering is assumed to be dominated by the mechanism illustrated in
Fig.\ref{f2dipole} in which the photon fluctuates into a $q\bar{q}$ pair of
fixed transverse separation $r$ and the quark carries a fraction $z$ of the
incoming photon light-cone energy. Using the Optical Theorem, this leads to
\begin{equation}
\sigma_{\gamma^{\ast}p}^{L,T}=\int dz\;d^{2}r\ |\Psi_{\gamma}^{L,T}%
(r,z,Q^{2})|^{2}\sigma(s^{\ast},r)\label{dipoledis}
\end{equation}
for the total virtual photon-proton cross-section, where $\Psi_{\gamma}^{L,T}$
are the appropriate spin-averaged light-cone wavefunctions of the photon and
$\sigma(s^{\ast},r)$ is the dipole cross-section. The dipole cross-section is
usually assumed to be independent of $z$, and is parameterised in terms of an
energy variable $s^{\ast}$ which depends upon the model.

Thus using Eq.(\ref{dipoledis}) we are able to compute the deep inelastic structure
function $F_2(x,Q^2)$. The power of the dipole model formulation lies in the fact that
the same dipole cross-section $\sigma(s^*,r)$ appears in a variety of other 
observables which involve the scattering of a real or virtual photon off a 
hadronic (or nuclear) target at high centre-of-mass (CM) energy. The largeness of the
CM energy guarantees the factorization of scattering amplitudes into a product of
wavefunctions and a universal dipole cross-section. In this paper we wish to
test the universality of the dipole cross-section using a wide range of high
quality data collected at the HERA $ep$ collider. Moreover, we also wish to examine
the extent to which to data are able to inform us of the role, if any, played by
non-linear saturation dynamics.

\section{The dipole cross-section}

We now turn to the three different models used to describe the dipole cross-section.
Before doing so however, we shall first discuss our choice of photon wavefunction.
For small $r$, the light-cone photon wavefunctions are given by the tree level
QED expressions~\cite{NZ91}.

These wavefunctions decay exponentially at large $r$, with typical $r$-values
of order $Q^{-1}$ at large $Q^{2}$ and of order $m_{f}^{-1}$ at $Q^{2}=0$.
However for large dipoles $r \approx 1$ fm, which are important at low $Q^{2}$,
a perturbative treatment is not really appropriate. In this region some
authors~\cite{FKS99} modify the perturbative wavefunction by an enhancement
factor motivated by generalised vector dominance (GVD)
ideas~\cite{GVD1,GVD2,FGS98}, while others~\cite{GW99a} achieve a similar but
broader enhancement by varying the quark mass\footnote{For a review of the
  colour dipole model, including a fuller discussion
of these points see Forshaw and Shaw \cite{FS06}.}. In practice \cite{FS04}, the difference
between these two approaches only becomes important when analysing the precise
real photoabsorption data from fixed-target experiments \cite{Caldwell78}.
Since we will not consider these data here, we will adopt the simpler practise
of using a perturbative wavefunction at all $r$-values, and adjusting the
quark mass to fit the data.

Turning now to the dipole cross-section, all three models are 
consistent with the physics of colour transparency for small dipoles and exhibit
soft hadronic behaviour for large dipoles. As stated above, the model parameters 
are determined by fitting only to the DIS structure function data.
Since the details of all three models have been published elsewhere, we shall 
here summarise their properties only rather briefly. 

\subsection{The FS04 Regge model \cite{FS04}}

This simple model, due to Forshaw and Shaw \cite{FS04},  combines colour transparency for small dipoles
$r<r_{0}$ with \textquotedblleft soft pomeron\textquotedblright\ behaviour for
large dipoles $r>r_{1}$ by assuming
\begin{align}
\sigma(x_{m},r) &  =A_{H}r^{2}x_{m}^{-\lambda_{H}}~~{\mathrm{for}}
~~r<r_{0}~~{\mathrm{and}}\nonumber\\
&  =A_{S}x_{m}^{-\lambda_{S}}~~{\mathrm{for}}~~r>r_{1},\label{eq:FS04}
\end{align}
where
$
x_{m}=(Q^{2} + 4m_f^{2}) /(Q^{2}+W^{2}) \, .
$
For light quark dipoles, the quark mass $m_f$ is a parameter in the fit, whilst
for charm quark dipoles the mass is fixed at 1.4~GeV. In the intermediate
region $r_{0}\leq r\leq r_{1}$, the dipole cross-section is given by
interpolating linearly between the two forms of Eq.(\ref{eq:FS04}).

If the boundary parameters $r_{0}$ and $r_{1}$ are kept constant then this
parameterisation reduces to a sum of two powers, as might be predicted in a
two pomeron approach, and can be thought of as an update of the original FKS
Regge model \cite{FKS99} to accommodate the latest data. It is plainly
unsaturated, in that the dipole cross-section obtained at small $r$-values and
fixed $Q^{2}$ grows rapidly with increasing $s$ (or equivalently with
decreasing $x$) without damping of any kind.

\subsection{The FS04 Saturation model \cite{FS04}}

Saturation can be introduced into the above model by adopting a method
previously utilised in \cite{MFGS2000}. Instead of taking $r_{0}$ to be
constant, it is fixed to be the value at which the hard component is some
specified fraction of the soft component, i.e.
\begin{equation}
\sigma(x_{m},r_{0})/\sigma(x_{m},r_{1})=f
\end{equation}
and $f$ instead of $r_{0}$ is treated as a parameter in the fit. This
introduces no new parameters compared to the Regge model. However, the scale
$r_{0}$ now moves to lower values as $x$ decreases, and the rapid growth of
the dipole cross-section at a fixed, small value of $r$ begins to be damped as
soon as $r_{0}$ becomes smaller than $r$. In this sense we model saturation,
albeit crudely, with $r_{0}$ the saturation radius.

\subsection{The CGC saturation model \cite{IIM04} }

In addition we shall consider the CGC dipole model originally presented by
Iancu, Itakura and Munier \cite{IIM04}. This model aims to include the main
features of the \textquotedblleft Colour Glass Condensate\textquotedblright%
\ regime, and can be thought of as a more sophisticated version of the
original \textquotedblleft Saturation Model\textquotedblright\ of
Golec-Biernat and W\"{u}sthoff \cite{GW99a}. The original Iancu et al
dipole cross-section was obtained using a three flavour fit to the DIS data
Here we  use a new four-flavour CGC fit due to  Kowalski, Motyka and Watt
\cite{KMW}. 

\section{Structure function data}

The parameters of the FS04 models were determined by fitting
the recent ZEUS $F_{2}$ data \cite{ZEUSf2data} in the kinematic range
\begin{equation}
0.045\,\mathrm{GeV}^{2}<Q^{2}<45\,\mathrm{GeV}^{2}\hspace{0.5cm}x\leq0.01\;
\end{equation}
whilst the CGC fit of \cite{KMW} is to data with $Q^2 > 0.25$ GeV$^2$ (the
other limits are as for FS04).
The corresponding H1 data \cite{H1f2data} could also be used, but it would
then be necessary to float the relative normalisation of the two data sets. We
do not do this since the ZEUS data alone suffice. The resulting parameter
values are tabulated in the original papers; we do not reproduce them here,
but confine ourselves to some general comments.

\begin{figure}[htb]
\begin{center}
\includegraphics*[width=7cm]{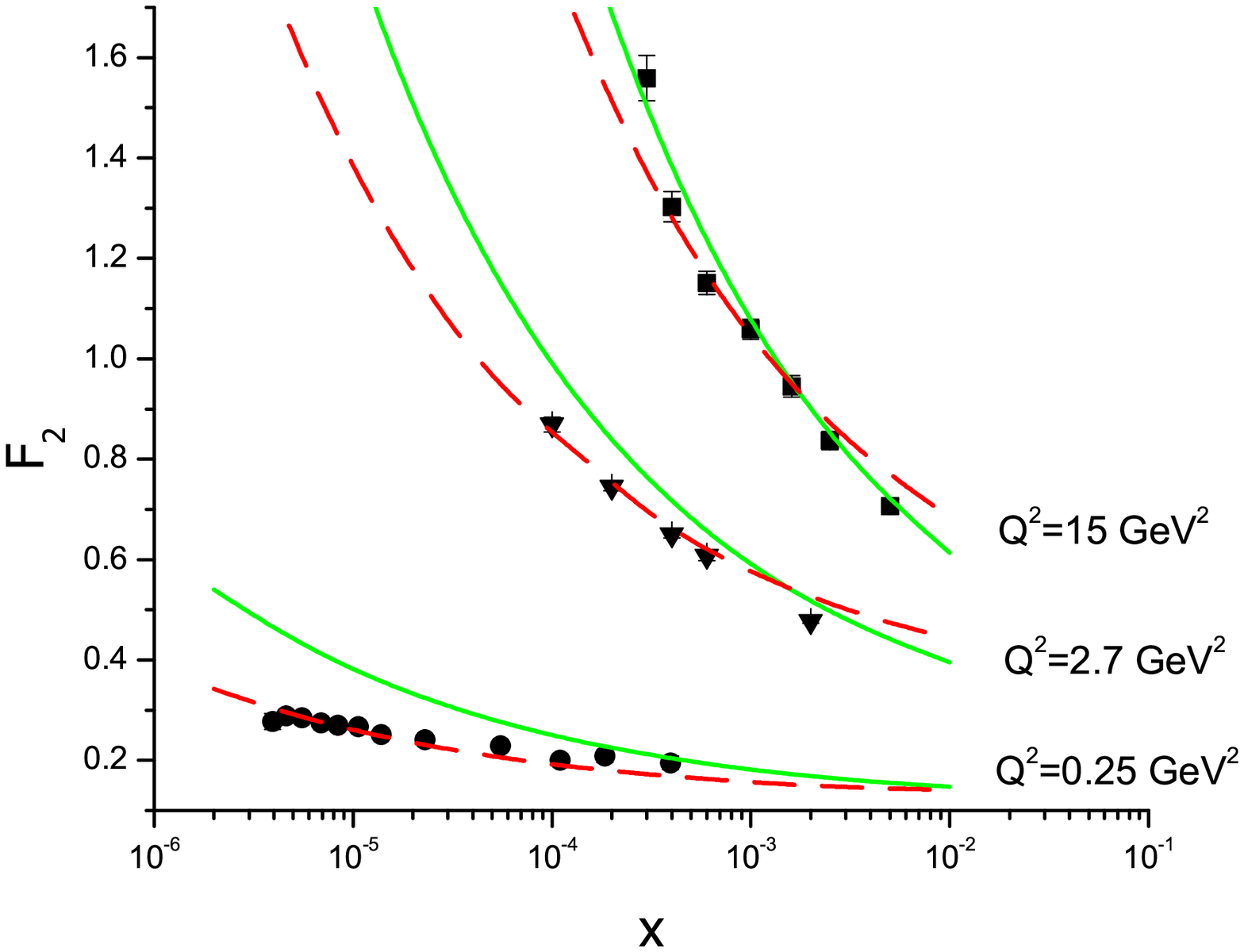}\includegraphics*[width=7cm]{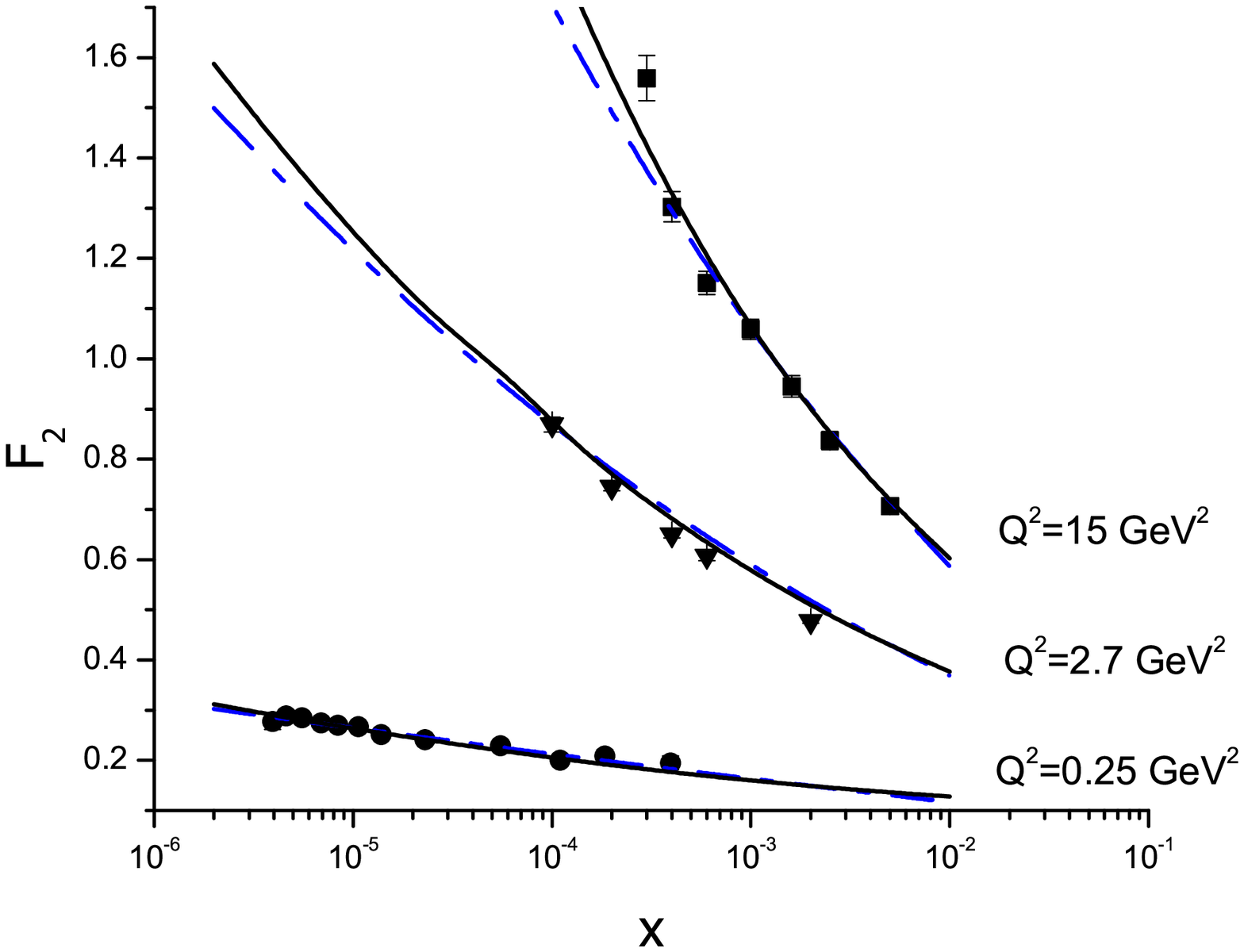}
\caption{Comparison of our new dipole model fits to a subset of DIS data \cite{ZEUSf2data}.
Left: No saturation fits. FS2004 Regge dipole fit (dashed line) and
(solid line) a fit 
of the same model to data in the restricted range $5 \times 10^{-4} < x <
10^{-2}$, extrapolated over the whole $x$-range $x < 0.01$.  
Right: Saturation fits. FS2004 saturation fit (solid line) and the CGC dipole
model (dot-dashed line)} 
\label{f2fits}
\end{center}
\end{figure}

The best fit obtained with the FS2004 Regge model 
is shown as the dashed line in Figure \ref{f2fits} (left). As can be seen, the
quality of the fit is not good, corresponding to a $\chi^2$/data point  of
428/156. This is not just a failing of 
this particular parameterisation. We have attempted to fit the data with many
other Regge 
inspired models, including the original FKS parameterizations
\cite{FKS99,MSS02}, without success.

A possible reason for this failing is suggested  by the solid curve in 
Figure \ref{f2fits} (left), which shows a
the result of fitting only to data in the $x$-range $5 \times 10^{-4} < x <
10^{-2} $, and then extrapolating the fit to lower $x$ values, corresponding to
higher energies at fixed $Q^2$. As can be seen, this leads to a much steeper
dependence at these lower $x$-values(i.e. at higher energies) than is allowed
by the data at all $Q^2$.  
An obvious way to solve this problem is by introducing saturation at high
energies, to dampen this rise. This is confirmed by fitting with the the FS04
saturation model, which yields  a $\chi^2$/data point of 155/156. The resulting
fit is shown in  Figure \ref{f2fits} (right), which also shows the very 
 similar results previously obtained using the more sophisticated CGC model.

It is clear from these results that
the introduction of saturation into the model immediately removes the tension between
the soft and hard components which is so disfavoured by the data. However, it is important
to note that this conclusion relies on the inclusion of the data in the low $Q^2$ region: 
both the Regge and saturation models yield satisfactory fits if we restrict to 
$ Q^2 \ge 2 \, {\rm GeV}^2$, with  $\chi^2$/data point values of 78/86 and 68/86
respectively.

At this point we have three well-determined parameterisations of the colour
dipole cross-section, which have been used by Forshaw, Sandapen and Shaw
\cite{fss:06a}   to yield predictions for other
processes. In the next sections we shall take a look at the results for  Deeply
Virtual Compton  
Scattering (DVCS)  and the diffractive structure function
$F_2^{D(3)}$\footnote{Predictions for vector meson
  production, which require a discussion of the vector meson wavefunctions,can
  be found in \cite{fss:04a,fss:06a}.} .We always choose to show the Regge fit,
even though it does not fit the 
$F_{2}$ data particularly well, in order to indicate the discriminatory power
of the data. We stress that in all cases, the photon wavefunctions and dipole
cross-sections are precisely those determined from the fits to $F_{2}$ data,
without any adjustment of parameters.

\section{Deeply virtual Compton scattering}

\begin{figure}[htb]
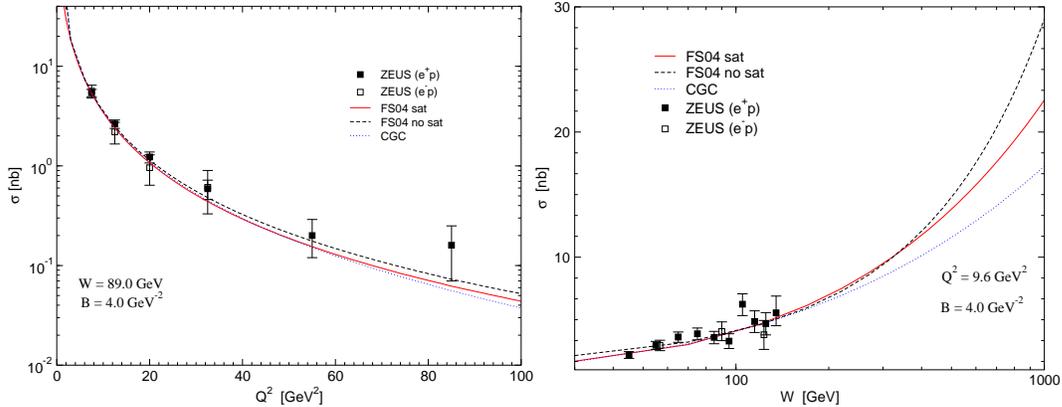

\begin{center}
\includegraphics*[width=7cm]{FIGS/DVCS_CGC_FS04_ZEUS_q2.eps}\includegraphics*[width=7cm]{FIGS/DVCS_CGC_FS04_ZEUS_W.eps}
\caption{Comparison of the ZEUS DVCS data \cite{zeusdvcs} with the predictions of the three
models discussed in the text.  Left : $Q^{2}$ dependence at $W=89$ GeV. Right:
 $W$ dependence at $Q^2 = 9.6$ GeV$^2$.}
\label{fig:DVCSZEUS}
\end{center}
\end{figure}

In deeply virtual Compton scattering, $
\gamma^{\ast}+p\rightarrow\gamma+p$, 
the final state particle is a real as opposed to a virtual photon and dipole
models provide 
predictions for the imaginary part of the forward amplitude with no adjustable
parameters beyond those used to describe DIS. To calculate the forward
cross-section a correction for the contribution of the real part of the 
amplitude has to be included. This correction was estimated in 
\cite{MSS02} and found to be less than $\approx 10\%$ and of a similar size 
in different dipole models. 
Predictions for the measured total cross-sections are then obtained 
using 
\begin{equation}
\sigma_{L,T}(\gamma^{\ast}p\rightarrow \gamma p)=\frac{1}{B}\left.  \frac
{d\sigma^{T,L}}{dt}\right\vert _{t=0}\;,\label{sigmatot}
\end{equation}
where the value of the slope parameter $B$ is taken from
experiment\footnote{For an alternative investigation of the link between
low-$x$ DIS  and DVCS and other exclusive proceses at high energy, see  Kuroda
and   Schildknecht \cite{KS}.}.

The predictions of all three dipole models are compared 
with the ZEUS data 
\cite{zeusdvcs} in Fig.\ref{fig:DVCSZEUS}\footnote{Note that throughout this paper the curves 
labelled `FS04 no sat' correspond to the predictions of the FS04 Regge model.},
where we take a fixed value $B=4$~GeV$^{-2}$ which is compatible with
their data. Bearing in mind this normalisation uncertainty, the agreement is good 
for all three models, although significant differences between the
models appear when the predictions are extrapolated to high enough energies,
as one would expect. Similarly good agreement is found for the H1 data
\cite{H105} (see \cite{fss:06a}). 

\section{Diffractive deep inelastic scattering (DDIS)}

\begin{figure}[h]
\begin{center}
\includegraphics[width=5cm]{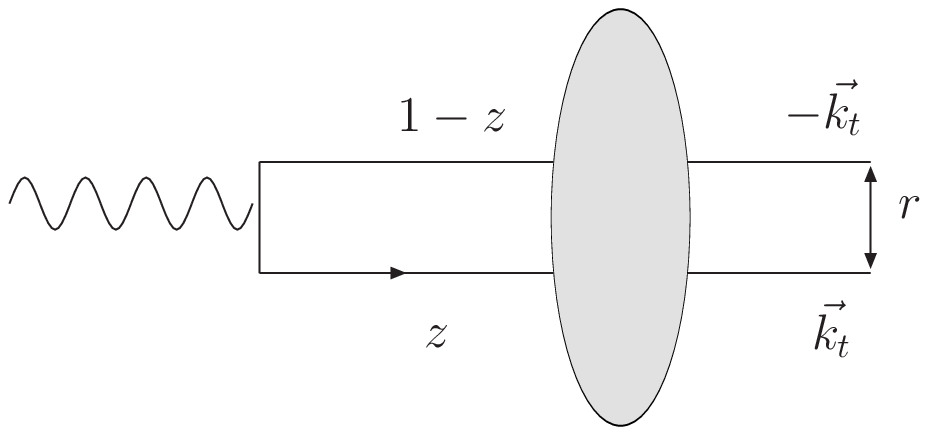}
\includegraphics[width=5cm]{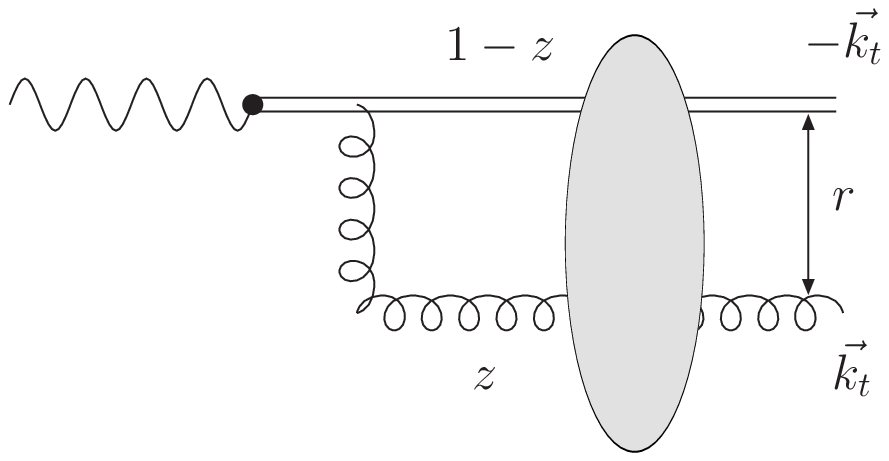}
\end{center}
\caption{The $q\bar{q}$ and $q\bar{q}g$ contributions to DDIS.}
\label{fig:feynman}
\end{figure}

To conclude our study we turn to the diffractive deep inelastic scattering
(DDIS) process
$$ \gamma^{\ast}+p\rightarrow X+p\;\;,\label{DDIS} $$
where the hadronic state $X$ is separated from the proton by a rapidity gap.
In this process, in addition to the usual variables $x$ and $Q^{2}$ there is a
third variable $M_{X}^{2}$. In practice,
$x$ and $M_{X}^{2}$ are often replaced by the variables $x_{I\!\!P}$ and
$\beta$:
\begin{equation}
x_{I\!\!P}\simeq\frac{M_{X}^{2}+Q^{2}}{W^{2}+Q^{2}}\hspace{1cm}\beta=\frac
{x}{x_{I\!\!P}}\simeq\frac{Q^{2}}{M_{X}^{2}+Q^{2}}~.
\label{eq:diffractive.variables}
\end{equation}
In the diffractive limit $s\gg Q^{2},m_{X}^{2}$ and so $x_{I\!\!P}\ll1$.

\begin{figure}[htb]
\begin{center}
\includegraphics*[width=15cm]{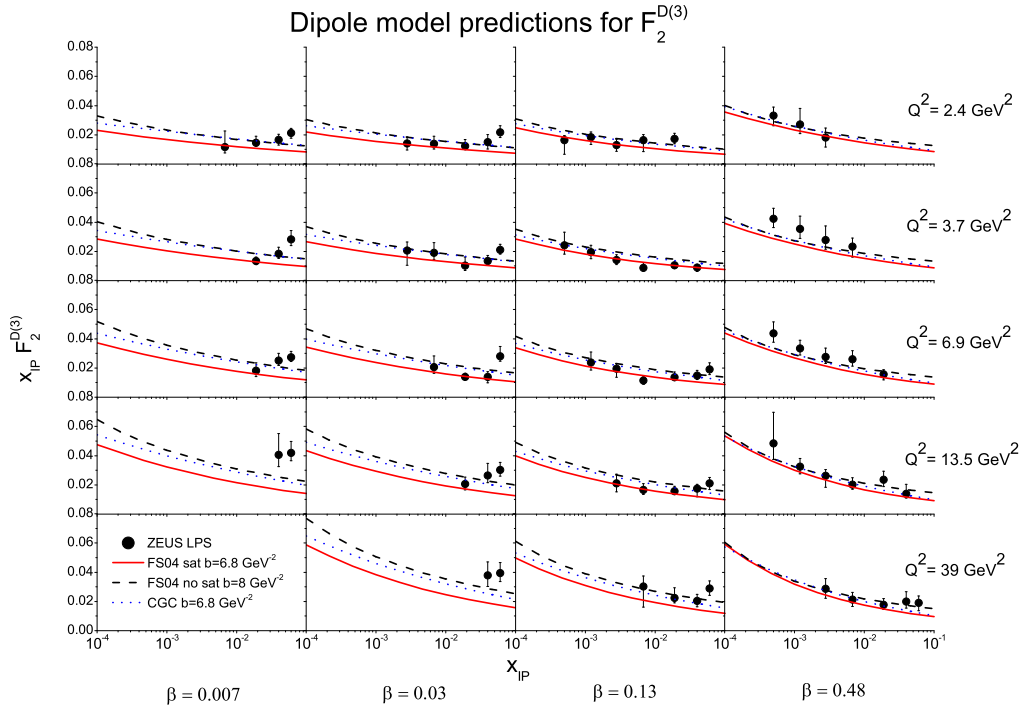}
\end{center}
\caption{Model predictions compared to the ZEUS LPS data \cite{LPS05}.}
\label{fig:F2D3LPS}
\end{figure}

In the dipole model, the contribution due to quark-antiquark dipoles to the
structure function $F_{2}^{D(3)}$ can be obtained from a momentum space
treatment as described in \cite{Wusthoff97,GW99a}. However, if we are to
confront the data at low values of $\beta$, corresponding to large invariant
masses $M_{X}$, it is necessary also to include a contribution from the higher
Fock state $q\bar{q}g$. We can estimate this contribution using an effective
\textquotedblleft two-gluon dipole\textquotedblright\ approximation due to
W\"{u}sthoff \cite{Wusthoff97}, as illustrated in Fig.\ref{fig:feynman}.

\begin{figure}[htb]
\begin{center}
\includegraphics*[width=15cm]{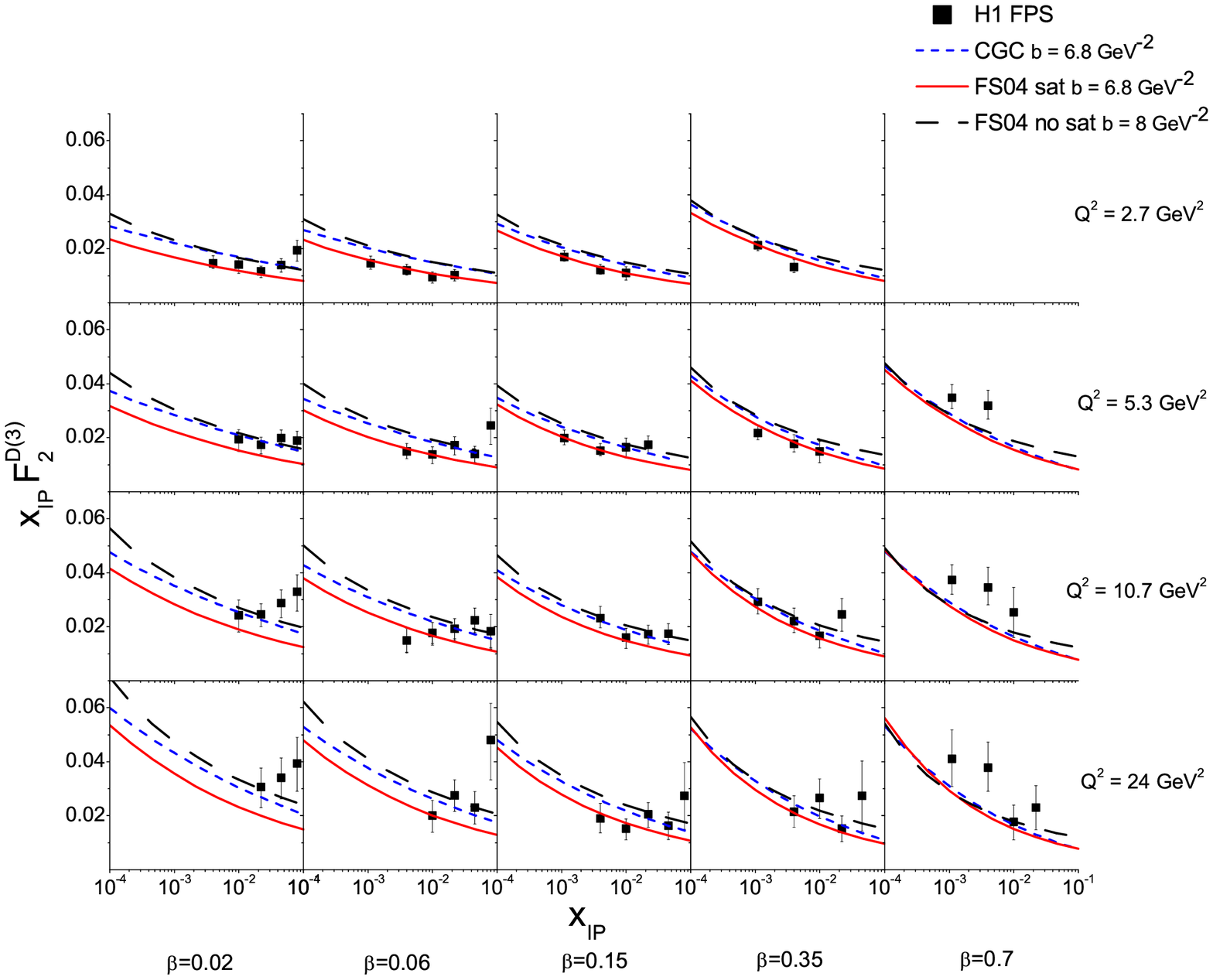}
\end{center}
\caption{Model predictions compared to the H1 FPS data: $\xp$ dependence \cite{H1FPS}.}
\label{fig:F2D3FPS1}
\end{figure}
\begin{figure}[htb]
\begin{center}\includegraphics*[width=13.5cm]{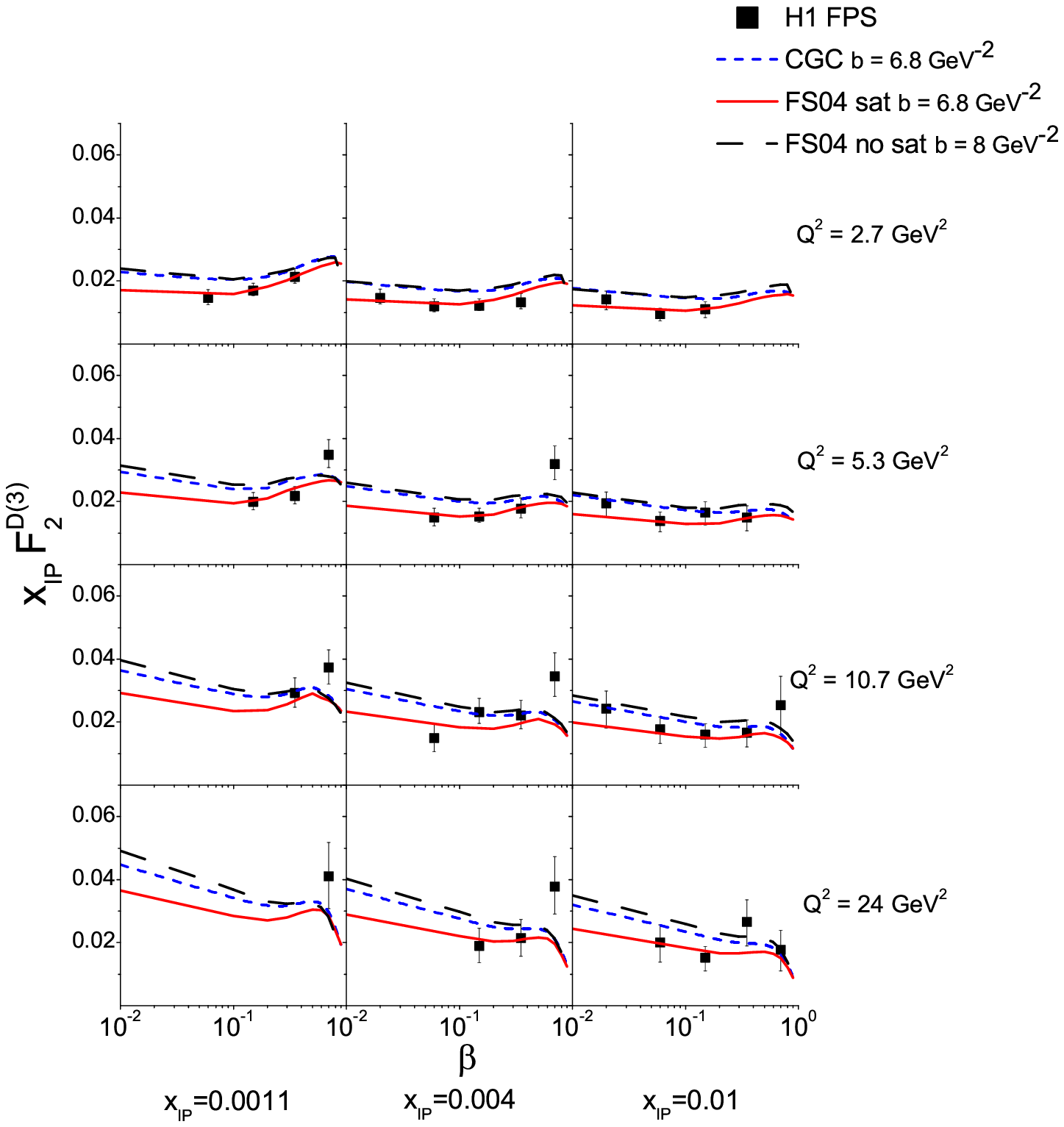}
\end{center}
\caption{Model predictions compared to the H1 FPS data: $\beta$ dependence \cite{H1FPS}.}
\label{fig:F2D3FPS2}
\end{figure}

Again, the predictions obtained in this way involve no adjustment of the
dipole cross-sections and photon wavefunctions used to describe the $F_{2}$
data. We are however free to adjust the forward slope for inclusive
diffraction ($B$) within the range acceptable to experiment, which means that
the overall normalisation, but not the energy dependence, of $F_{2}^{D(3)}$ is
free to vary somewhat. We take $B=6.8$ GeV$^{-2}$ when making our 
CGC and FS04 saturation predictions and $B=8.0$ GeV$^{-2}$ when 
making the FS04 Regge predictions. Note that a value of $8.0$ GeV$^{-2}$ is rather
high compared to the $\approx 6$ GeV$^{-2}$ favoured by the H1 FPS data \cite{H1FPS}
although it is in the range allowed by the ZEUS LPS data \cite{LPS05}. 
The need for a larger value of $B$ for the FS04 Regge model arises since the
corresponding dipole cross-section is significantly larger than the FS04 saturation
model at large values of $r$ and this enhancment is magnified in inclusive 
diffraction since it is sensitive to the square of the dipole cross-section.
We should also bear in mind that the tagged proton data are subject to an overall 
$\approx 10\%$ normalisation uncertainty. We are also somewhat free to vary 
the value of $\alpha_{s}$ used to define the normalisation of the model 
dependent $q\bar{q}g$ component, which is important at low values of $\beta$. Rather
arbitrarily we take $\alpha_{s}=0.1$ and take the view that the theory curves
are less certain in the low $\beta$ region.

In Fig.\ref{fig:F2D3LPS} we compare the recent ZEUS LPS data \cite{LPS05} on
the $x_{I\!\!P}$ dependence of the structure function $F_{2}^{D(3)}$ at
various fixed $Q^{2}$ and $\beta$ with the models\footnote{Predictions for the
original Iancu et al CGC model have previously been published in
\cite{fss:04b}.}. The agreement is good except at the larger $x_{I\!\!P}$
values. Indeed, the $\chi^2$ values per
data point are very close to unity for all three models for $\xp < 0.01$. 
Disagreement at larger $\xp$ is to be expected since this is the 
region where we anticipate a significant non-diffractive contribution which is absent 
in the dipole model prediction. Note that the three models produce similar 
predictions at larger values of $\beta$.

Comparison to the H1 data with tagged protons \cite{H1FPS} is to be found in 
Fig.\ref{fig:F2D3FPS1} and Fig.\ref{fig:F2D3FPS2}. The story is similar to that for
the ZEUS data and the evidence for an overshoot of the CGC and Regge
model predictions at low $\beta$ is strengthened. The $Q^2=2.7$ GeV$^2$ panes 
in Fig.\ref{fig:F2D3FPS2} illustrate this point the best. Again, we should
not interpret this as evidence against these dipole models due to the uncertainty in
the $q\bar{q}g$ contribution in the low $\beta$ region. The agreement between all
models and the data at larger values of $\beta$ and low enough $\xp$
is satisfactory.\footnote{A fuller discussion, including comparison with the
  ZEUS FPC data \cite{FPC05} and the H1 M$_{\rm Y}$ data \cite{H1MY} is given in \cite{fss:06a}.}.

In summary, the DDIS data at large enough $\beta \ge 0.4$ and small enough 
$\xp \le 0.01$ are consistent with the predictions of all three dipole 
models. However the data themselves would have a much greater power to
discriminate between models if the forward slope parameter were measured to
better accuracy. At smaller values of $\beta$, the data clearly reveal the 
presence of higher mass diffractive states which can be estimated via the
inclusion of a $q\bar{q}g$ component in the dipole model calculation
under the assumption that the three-parton system interacts as a single dipole
according to the universal dipole cross-section. The theoretical calculation
at low $\beta$ must be improved before the data in the region can be utilised 
to disentangle the physics of the dipole cross-section. Nevertheless, it is
re-assuring to observe the broad agreement between theory and data in the low
$\beta$ region.    

\section{Conclusion}
The dipole scattering approach, when applied to diffractive 
electroproduction processes, clearly works very well indeed. The HERA
data now constitute a large body of data which is typically accurate to 
the 10\% level or better, and without exception the dipole model is able
to explain the data in terms of a single universal dipole scattering
cross-section. Perhaps the most important question to ask of the data is
the extent to which saturation dynamics is present. Although the $F_2$ 
data suggest the presence of saturation dynamics \cite{FS04}, 
the remaining data on exclusive processes and on $F_2^{D(3)}$ 
are unable to distinguish between the models we consider here:
these data are therefore unable to offer additional information on the possible
role of saturation. We do note that a more accurate determination of the forward
slope parameter in diffractive photo/electro-production processes would 
significantly enhance the impact of the data. However, it is hard to avoid the
conclusion that only with more precise data or with data out to larger values of 
the centre-of-mass energy will we have the chance to make a definitive statement on 
the role of saturation without the inclusion of the low $Q^2$ $F_2(x,Q^2)$ data in
the analysis.

\end{document}